\def\K{{\rm\thinspace K}}

\def\Msun{\hbox{$\rm\thinspace M_{\odot}$}}

\def\s{{\rm\thinspace s}}

\documentclass{iau}
\usepackage{graphicx}

\title[Probing GR with Black Holes] 
{Probing General Relativity with Accreting Black Holes}

\author[A.C. Fabian]   
{A.C. Fabian $^1$}

\affiliation{$^1$ Institute of Astronomy,\\ Madingley Road.\\Cambridge
  CB3 0HA\\UK \\ email: {\tt acf@ast.cam.ac.uk} \\[\affilskip]}

\pubyear{2012}
\volume{290}  
\jname{Feeding compact objects: Accretion on all scales}
\editors{C.M. Zhang, T. Belloni, M. M\'endez \& S.N. Zhang, eds.}

\begin{document}

\maketitle

\begin{abstract}Most of the X-ray emission from luminous accreting
  black holes emerges from within 20 gravitational radii. The
  effective emission radius is several times smaller if the black
  hole is rapidly spinning. General Relativistic effects can
  then be very important. Large spacetime curvature causes strong
  lightbending and large gravitational redshifts. The hard X-ray,
  power-law-emitting corona irradiates the accretion disc generating an
  X-ray reflection component. Atomic features in
  the reflection spectrum allow gravitational redshifts to be
  measured.  Time delays between observed variations in the power-law
  and the reflection spectrum (reverberation) enable the physical
  scale of the reflecting region to be determined. The relative
  strength of the reflection and power-law continuum depends on light
  bending. All of these observed effects enable the immediate 
  environment of the black hole where the effects of General
  Relativity are on display to be probed and explored.

\end{abstract}

\firstsection 

\section{Introduction}
Black holes are a common feature of the Universe. They have long been
suspected to be responsible for the prodigious powers of quasars
(Lynden-Bell 1969) and Galactic X-ray binaries such as Cygnus X-1
(Tananbaum et al 1972). In these cases it is accretion of matter into
the black hole which makes these objects luminous. We do not of course
``see'' the black hole here, but the matter swirling around it in the
accretion flow which is heated by the gravitational energy
released. This process is the most efficient known, in terms of the
fraction of rest mass released, after matter-antimatter annihilation,
with a typical value of ten per cent, which is about 20 times higher
than hydrogen fusing into helium.

Much of the energy is released in the strong gravity regime very close
to the black hole at only a few gravitational radii ($r_{\rm
  g}=GM/c^2$) enabling the effects of general relativity to be probed
by observation. Strongly curved spacetime leads to large gravitational
redshifts, strong light bending and if the black hole is spinning,
dragging of inertial frames, revealed through the Innermost Stable
Circular Orbit or ISCO. We concentrate here on luminous accreting
black holes and in particular on their X-radiation. The observed rapid
variability seen in some Active Galactic Nuclei (AGN) has long shown
that the X-ray emission originates from a physically small
region. Relativistically blurred reflection components and
reverberation now show just how small a region this can be.

\begin{figure}
\begin{center}
\includegraphics[angle=0,width=2.4in]{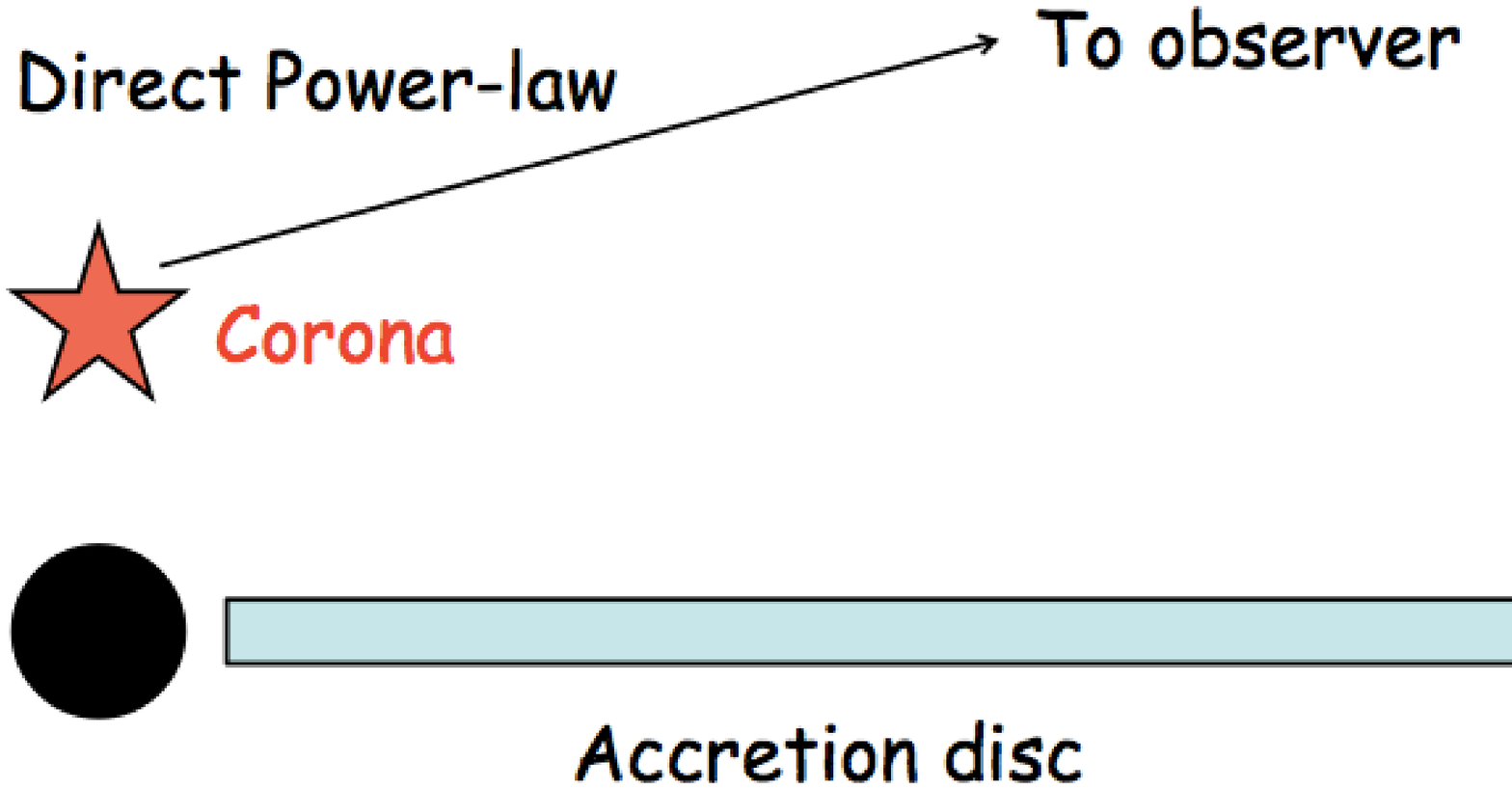}
\hspace{1cm}
\includegraphics[angle=0,width=2.4in]{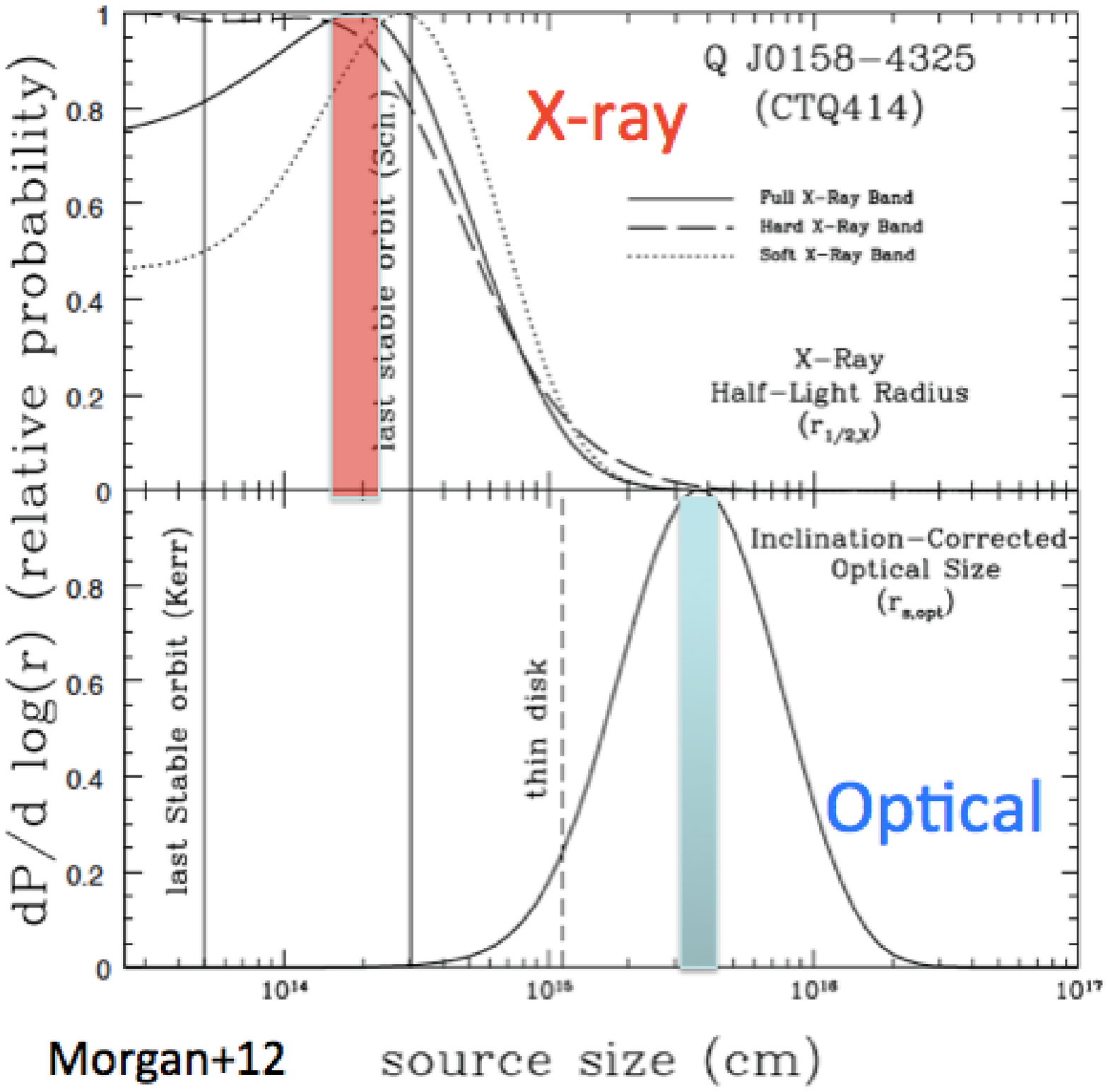}
\end{center}
\caption{Left: The primary emission components of a luminous accreting
  black hole consist of the power-law continuum emitting corona and
  the quasi-black body disc. Right: Estimates of the half-light radii
  of the X-ray (upper) and optical (lower) emitting regions of the
  doubly-imaged lensed quasar Q\,0158-4325 obtained from optical and
  X-ray monitoring of its rapid microlensing variability (Morgan et al
  2012).  }
\end{figure}

\section{X-ray Reflection }

Matter accreting onto a black hole is most unlikely to fall in
radially but will have sufficient angular momentum to go into orbit
about it. Viscosity then causes the matter to spiral inward while the
angular momentum is transferred outward. This forms an accretion disc
which will be dense, optically thick and physically thin provided that
the gravitational energy released is radiated locally. The emitted
spectrum is a quasi-blackbody of temperature of about $10^7\K$ for a
luminous disc around a stellar mass black hole and drops to about
$10^5\K$ for a billion solar mass black hole. The accreting gas will
therefore be a hot dense plasma with the differential rotation winding
up the magnetic fields (which provides the viscosity). In a manner
similar to the production of coronal magnetic structures on and above
the Sun, we can expect that magnetic structures will occur above the
inner accretion disc. Magnetic reconnection in this corona can
accelerate particles in the corona and, through inverse Compton
scattering of soft disc photons, produce a hard power-law continuum.

Such a configuration (Fig.~1, left) explains the basic spectral
components of an Active Galactic Nucleus (AGN); a big blue bump of
quasi-blackbody emission from the disc itself and a hard power-law of
X-rays extending to hundreds of keV. The coronal power-law emission
can be rapidly variable due to its magnetic nature. The rapid X-ray
variability seen in many sources shows that the power-law source in
most accreting black holes is compact in size. Clear evidence of this
compactness is provided by microlensing studies of quasars which are
multiply imaged by an intervening galaxy (e.g. Morgan et al 2012). The
half-light radius for the X-ray emission for Q\,0158-4325, shown in
Fig.~1, is less than $6r_{\rm g}$.

\begin{figure}
\begin{center}
\includegraphics[angle=0,width=4in]{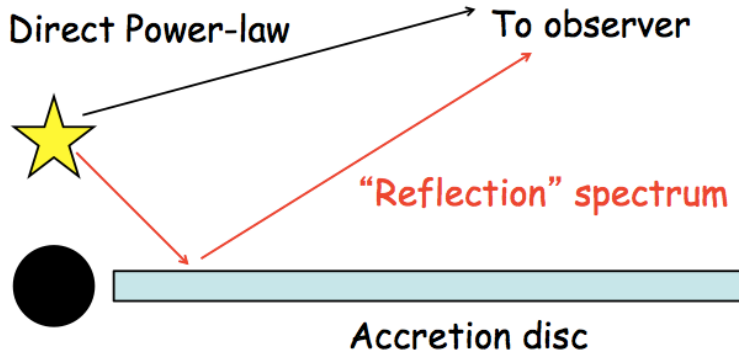}
\end{center}
\caption{Power-law X-ray source, indicated by the yellow blob
  representing the corona, irradiating the inner regions of the
  accretion disc about a black hole. The paths of the primary and
  reflected components are shown.  }
\end{figure}

The irradiation of the dense disc by the coronal power-law continuum
provides a further spectral component, X-ray reflection (Fig.~2). This
is just the back-scattered emission plus fluorescence, recombination
and bremsstrahlung (see Fabian \& Ross 2011 for a review). It consists
of a hard Compton hump together with a soft excess of re-emission
including emission lines. The strongest such line is usually iron
K$\alpha$ at 6.4-6.95~keV, depending on the ionization state of iron
in the disc. It is likely that the irradiation will be intense enough
to control and raise the ionization state above that expected from
the hot disc alone.

\begin{figure}
\begin{center}
\includegraphics[width=2.5in]{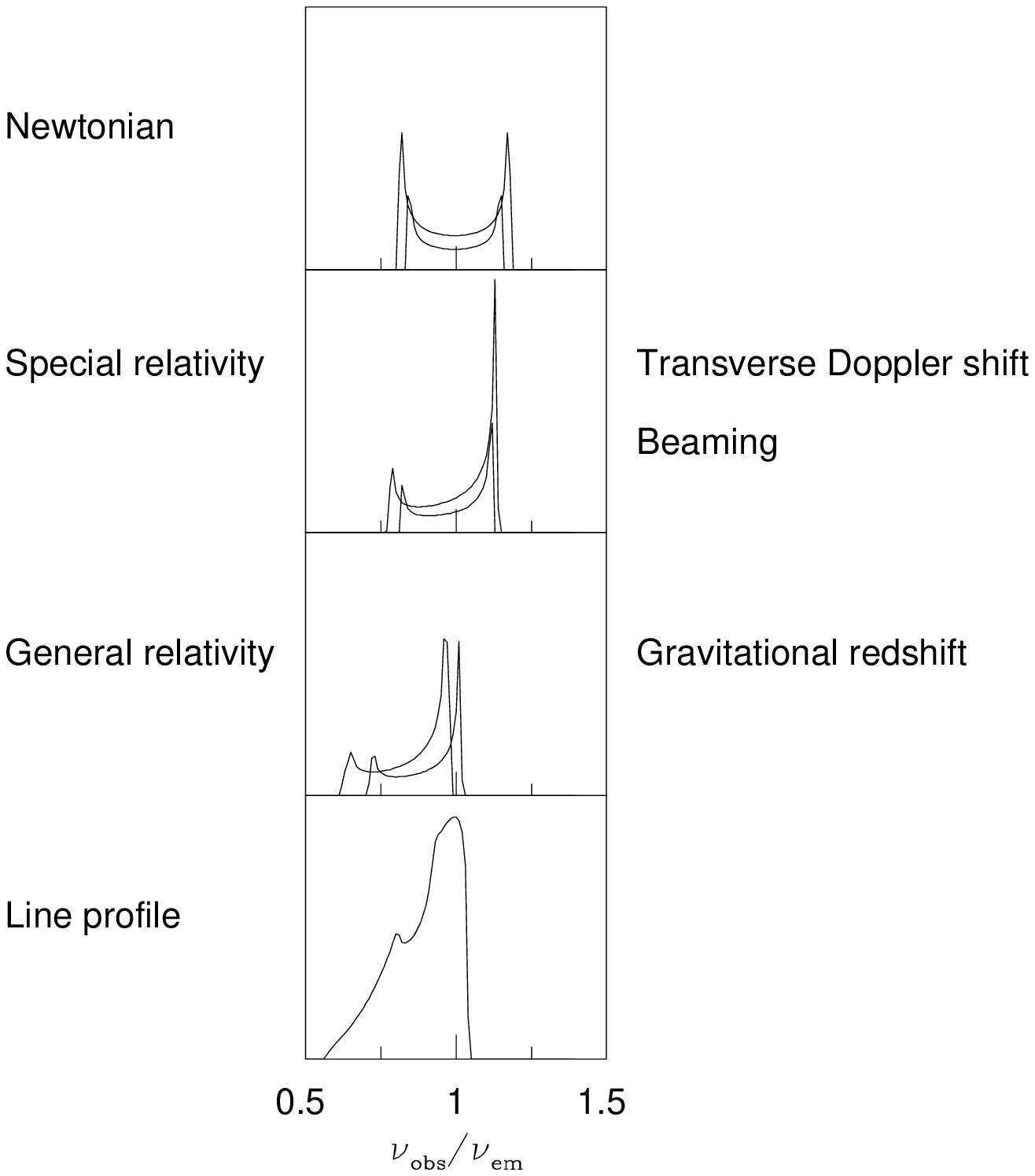}
\raisebox{1.5cm}{\includegraphics[width=1.7in]{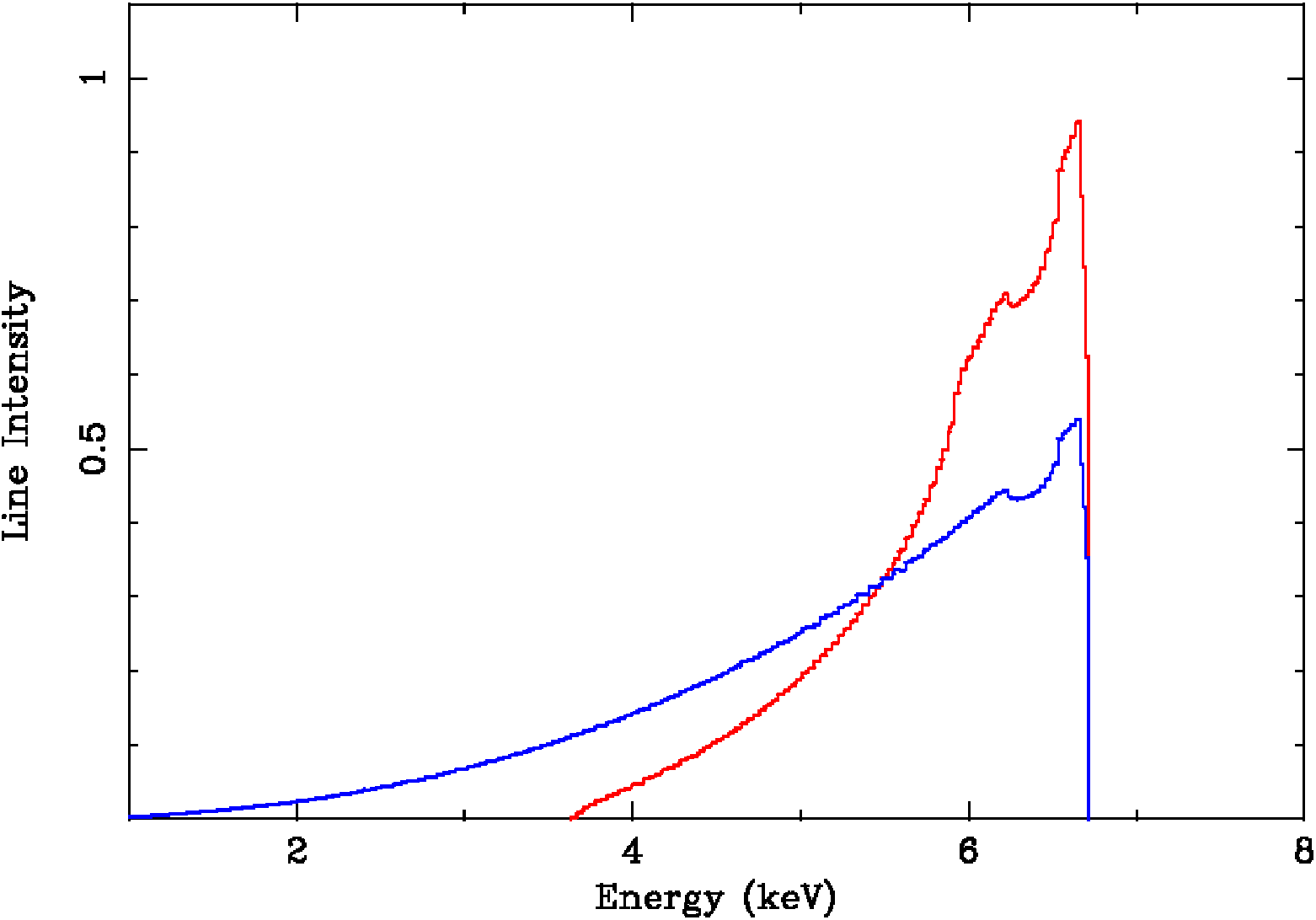}}
\end{center}
\caption{Left: Expected line profiles from 2 radii in an orbiting disc showing
  Newtonian, Special and General Relativistic effects. The lower panel
shows the broad skewed line expected from the whole disc. Right:
Broad iron line expected from a non-spinning Schwarzschild black hole
(red) and a maximally spinning Kerr black hole (blue).}
\end{figure}

\begin{figure}
\begin{center}
\includegraphics[width=3.5in,angle=0]{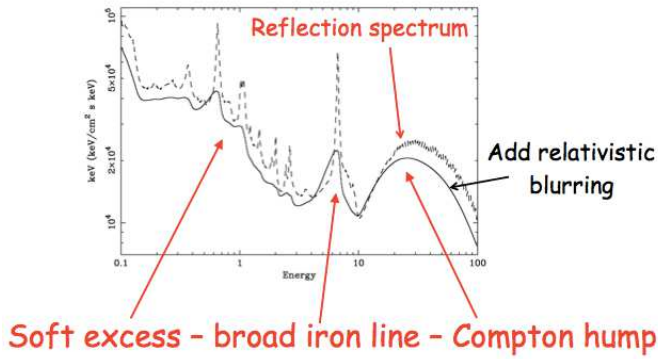}
\end{center}
\caption{  Relativistically-blurred reflection spectrum (unblurred shown by
  dashes) from Ross \& Fabian (2005). }
\end{figure}

\begin{figure}
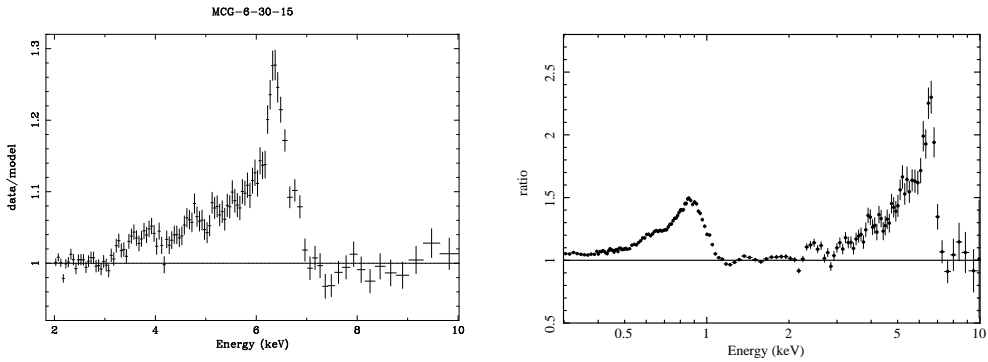

\begin{center}
\includegraphics[width=1.8in,angle=-90]{mcg6_linebw.ps}
\hspace{0.5cm}
{\raisebox{-0.35cm}{\includegraphics[width=1.7in,angle=-90]{2laor_rat.ps}}}
\end{center}
\caption{Left: Broad iron-K line in MCG--6-30-15 as seen with Suzaku
  (Miniutti et al 2007). The large red wing, extending to lower
  energies, is principally due to the effects of gravitational
  redshifts close to the black hole. Right: Broad iron-K and iron-L
  lines in 1H0707-495 (Fabian et al 2009). The small bump in the red
  wing of the iron-L line could be due to an oxygen-K line.}
\end{figure}

Having emission lines produced from the disc is very important, since
if they are observed then we can measure the Doppler shifts and thus the
velocity of the accretion flow, which can be up to half the speed of
light. We can also measure the gravitational redshift, which can tell
us the radius at which the emission originates (Fig.~6, Fabian et al
1989; Laor 1991). We expect an accretion disc to extend down to the
ISCO within which the matter plunges on a ballistic orbit into the
black hole. Since the radius of the ISCO depends on the spin of the
black hole, measurement of the largest gravitational redshift
translates to a measurement, or at least a lower limit on, the black
hole spin.

\begin{figure}
\begin{center}
\includegraphics[width=2.5in]{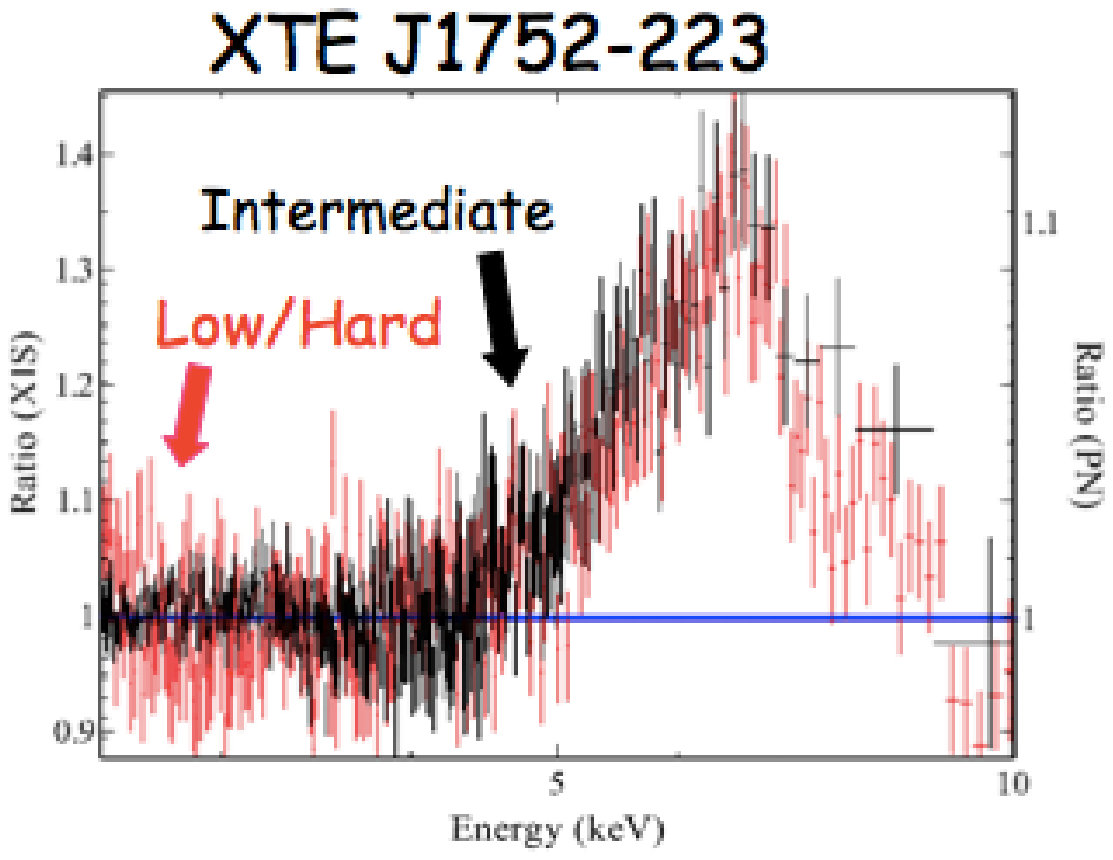}
\hspace{1cm}
{\includegraphics[width=1.7in]{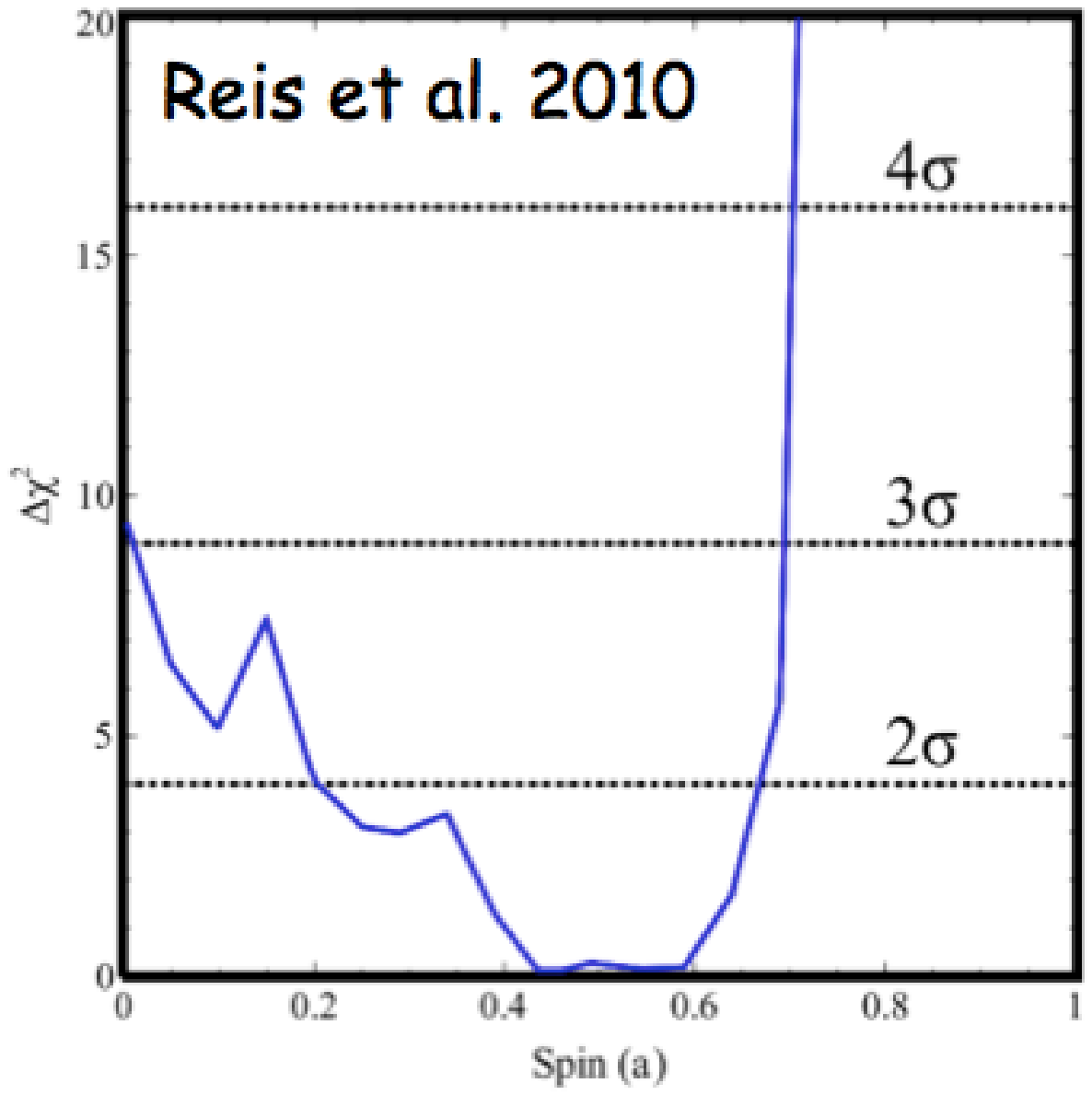}}
\end{center}
\caption{Left: Spectrum of the broad iron line in BHB XTE\,J1752-223
  (Reis et al 2011). The line profile is unchanged with spectral
  state. Spectral fitting shows that the black hole spin is about
  0.55. }
\end{figure}

The net observed spectrum from the inner parts of an accretion disc
around a black hole therefore consists of a power-law continuum, a
reflection spectrum blurred by Doppler shifts and gravitational
redshifts, a soft excess below 2~keV, a broad iron line from 4-7~keV
and a Compton hump peaking around 30~keV (Fig.~4). The broad iron line was
first seen from an AGN (MCG--6-30-15) with the Japanese-US satellite
mission, ASCA (Tanaka et al 1995).  A recent version of the spectrum
from this object using Suzaku is shown in Fig.~5 (Miniutti et al
2007). The line is so broad that indicates a high spin ($a>0.95$),
with the ISCO well within radius $r=2r_{\rm g}$, where GR effects must
be very strong.  Broad iron lines have been found from a range of
Seyfert 1 AGN (Nandra et al 2007, Brenneman \& Reynolds 2009),
Galactic Black Hole Binaries (BHB, Miller 2007) and neutron star
systems (Cackett et al 2008). An example from a BHB is shown in Fig.~6
where the broad line of XTE\,J1752-223 appears not to change between
the intermediate and hard state during its 20?  outburst (Reis et
al 2010). Fitting the spectrum with a reflection model reveals the
black hole spin to be about 0.55.

The discussion has so far ignored absorption and outflows (e.g. fast
and slow warm absorbers) which clearly occur in some luminous
accreting black holes. Some (e.g. Miller et al 2008) argue that a
dense outflow completely obscures all emission from within tens to
hundreds $r_{\rm g}$, hides the strong gravity regime from view. The
absorber then mimics the relativistically blurred reflection features
described above. How such mimicry works for both AGN and BHB is not
explained.  Many BHB are bright enough that any such extreme
obscuration would be obvious. It is far more likely that absorption
and outflows are just a complication that can be corrected for in
spectral models. Microlensing, rapid X-ray variability and
reverberation results all point to the strong gravity regime being
directly observable in most Type~I AGN.

\section{Light Bending}

The iron line in MCG--6-30-15 is stronger than expected and, although
variable, does not show the same level of variability as the power-law
continuum (Fabian \& Vaughan 2003). These are likely to be due to
strong light bending (Strength: Martocchia \& Matt 1996, 2002;
Variability: Miniutti \& Fabian 2004). Strong gravity close to the
black hole bends the light towards it, causing it to be focussed on
the disc. In so doing it  makes an intrisically isotropic continuum source appear
anistropic to the outside observer (Fig.~7). Fewer photons escape to
infinity and more strike the disc, or fall into the hole as the corona
lies closer to the black hole. Good evidence for light bending was
noted in the BHB XTE\,J1650-300 by Rossi et al (2005). A recent
reworking of the data on that source by Reis et al (2012) shows this
result more clearly (Fig.~7).

\begin{figure}
\begin{center}
\includegraphics[width=2.4in,angle=0]{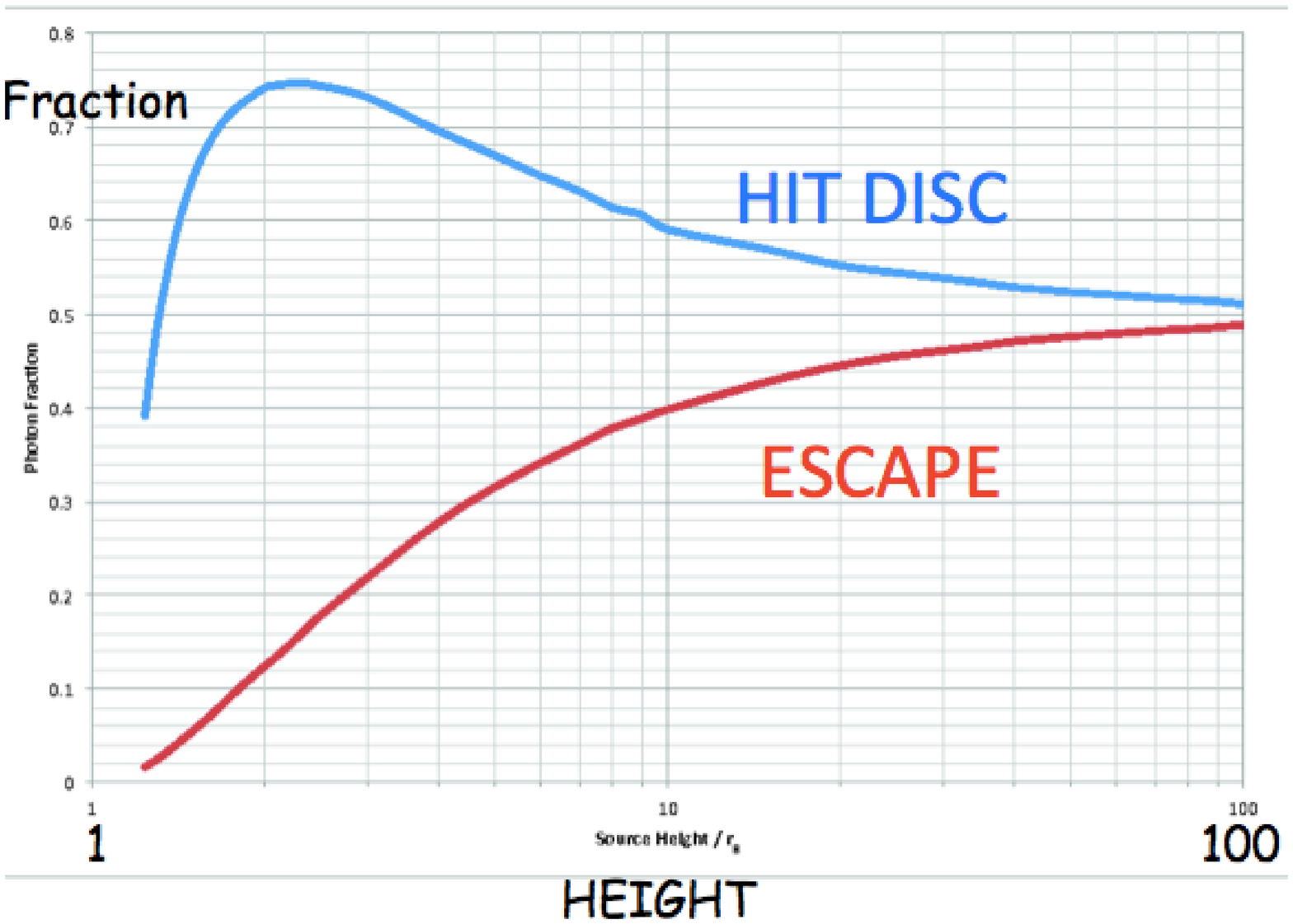}
\hspace{0.7cm}
\includegraphics[width=2.4in,angle=0]{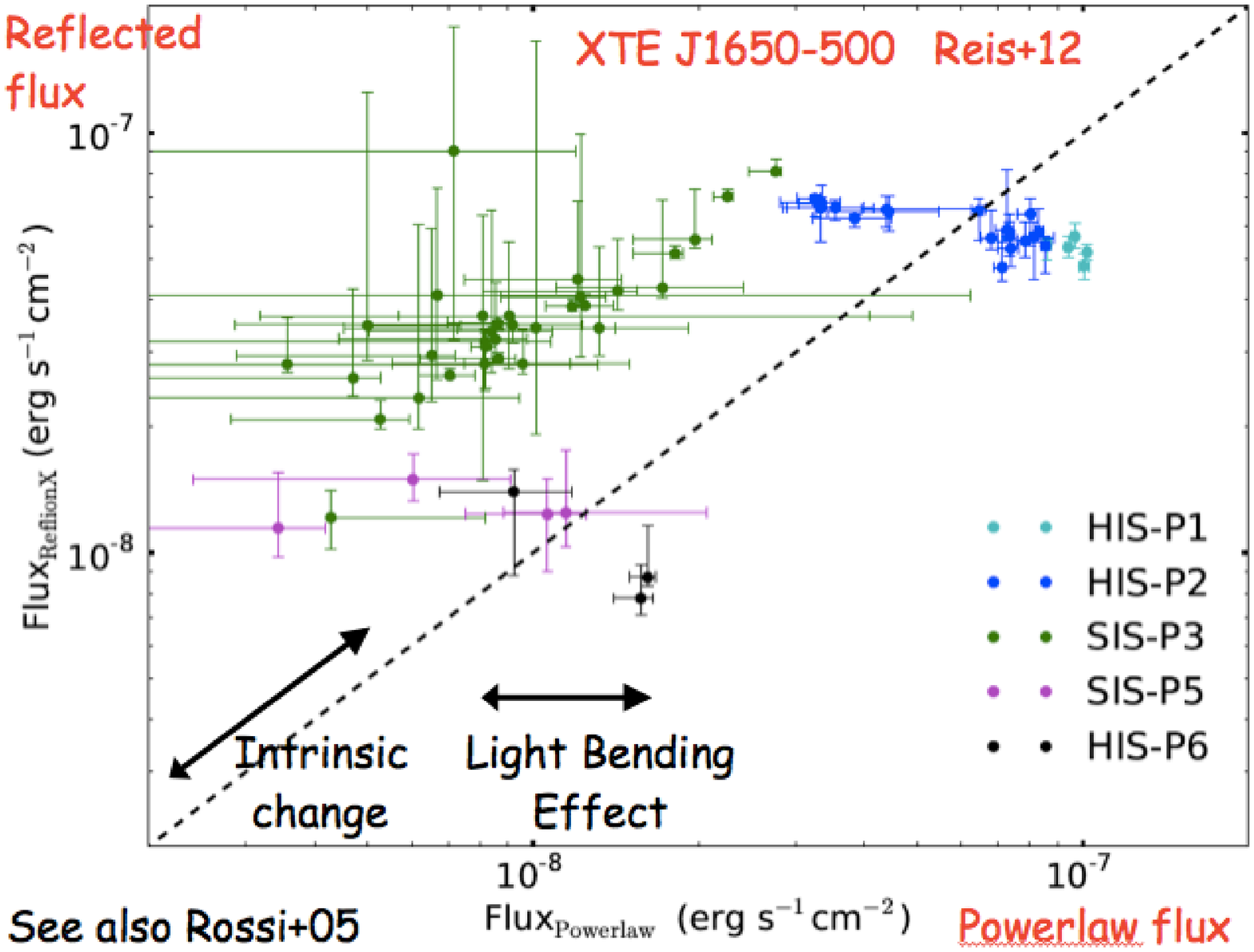}
\end{center}
\caption{Left: Fraction of photons from an isotropic sources which hit
  the disc (top blue line) or escape to infinity (lower red line) as a
  function of source height. Right: Intrinsic power-law flux plotted
  against flux in the reflection component in the outburst of
  the Galactic BHB XTE\,J1650-500 (Reis et al 2012). Intrinsic source
  changes cause motion along a diagonal of positive gradient, changes
  of coronal height cause motion along a diagonal of negative to zero
  gradient. The different coloured points indicate different states of
  the source. The evolution of the source in the plot, starting from
  the upper right, is interpreted as follows: the leftward shift of the
  points during the High Intermediate State (HIS) is due to the 
  corona dropping in height, it then weakens in intrinsic luminosity
  during the Soft Intermediate State (SIS) before rising back in
  height et the end of the SIS.}
\end{figure}

\begin{figure}[h]
\begin{center}
\includegraphics[width=2.6in]{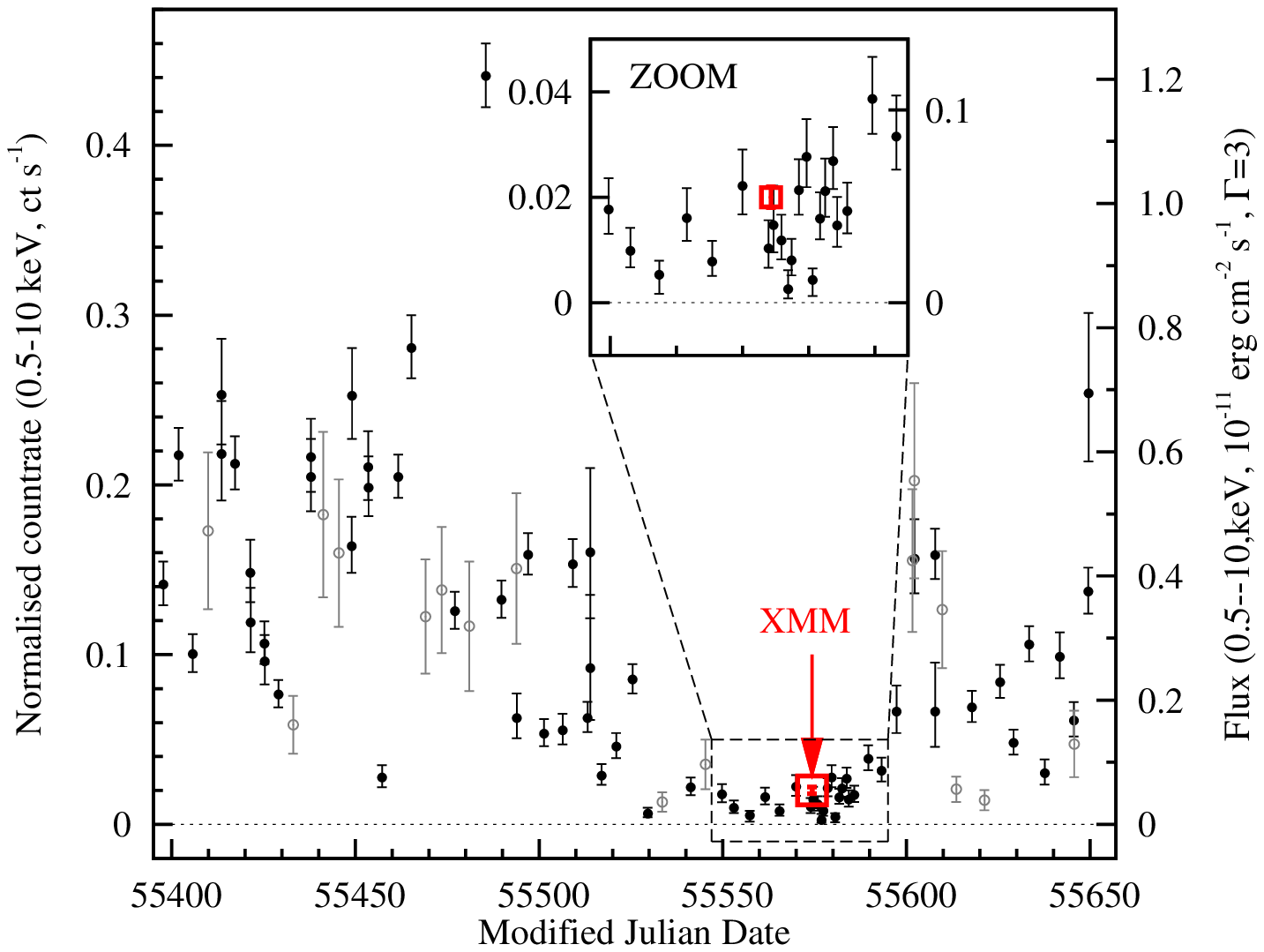}
\raisebox{0.5cm}{\includegraphics[width=2.3in]{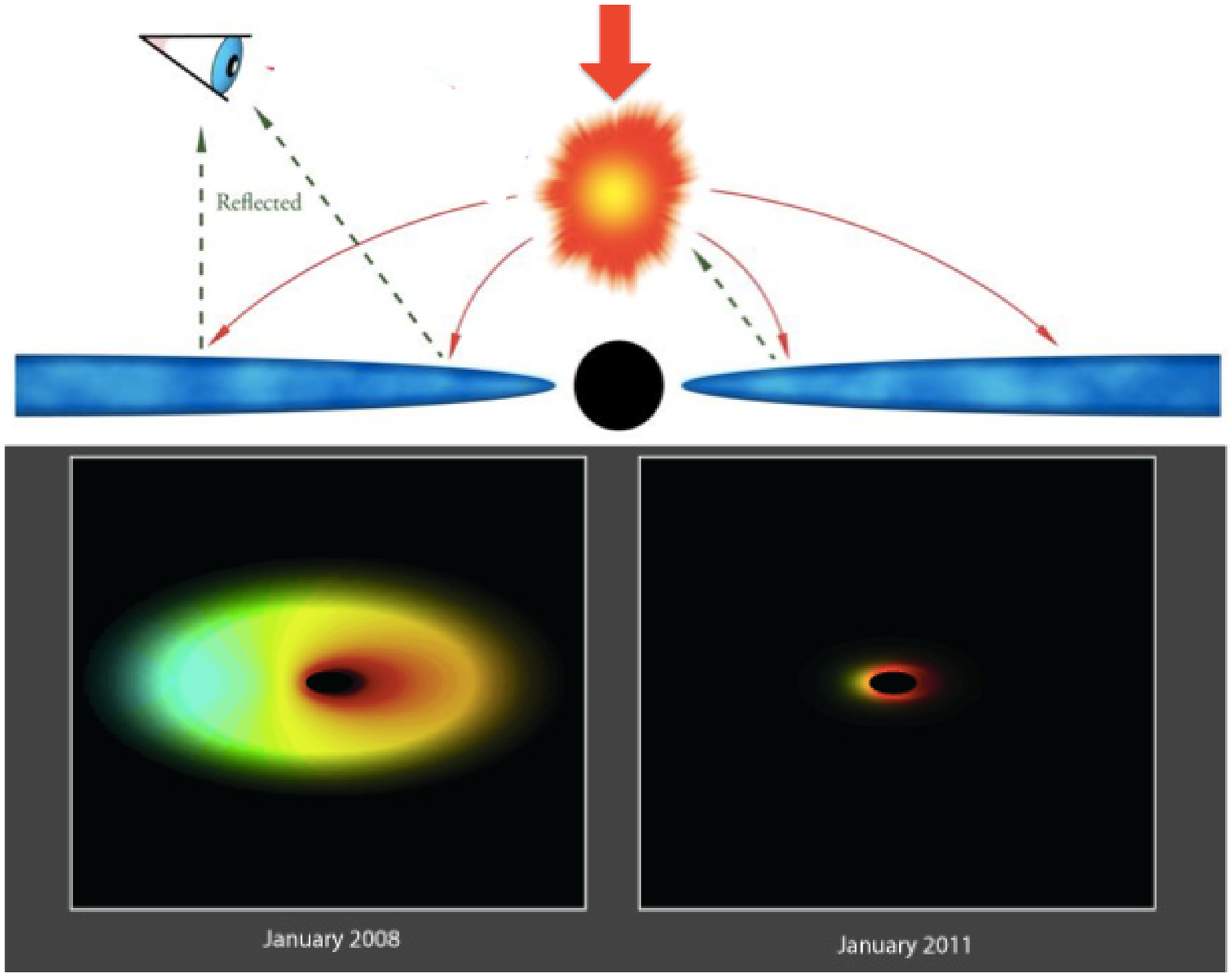}}
\end{center}
\caption{Top: Emissivity profile of 1H\,0707-495 in 2008 and 2011
  (Fabian et al 2012). The outer parts of the corona disappeared in
  early 2011
  (bottom panel, courtesy Dan Wilkins).}
\end{figure}

Further evidence in support of extreme light bending emerged from
early in 2011 when colleagues discovered that 1H0707-495 was
dramatically reducing in flux and going into a low state (Fig.~8). The
soft flux from the source reduced by over an order of magnitude in
January and February before recovering in March. We triggered an
observation of the object with XMM, under an accepted programme of
Norbert Schartel for studying low states in AGN. The spectrum of the
source now looked similar in shape to when brighter, although of
course much reduced in flux and most interestingly shifted to lower
energies. The corona had dropped to within
$2r_{\rm g}$ of the black hole (Fig.~8).  The shape of the emissivity
profile of reflection (deduced from the shape of the broad iron line)
required the very strong gravitational light bending expected very
close to the black hole (Wilkins \& Fabian 2011, 2012a).

Several other AGN have been seen to drop to a reflection-dominated
phase, explainable by extreme light bending due to the corona
collapsing to the centre (e.g. PG2112, Schartel et al 2007; Mkn\,335,
Gallo et al 2012; PHL\,1092, Miniutti et al 2012).

\section{Reverberation}

Recently, we have seen both the iron K$\alpha$ and L$\alpha$ lines in
the AGN 1H\,0707-495 (Fig.~5). This object is a very highly variable
type of AGN known as a Narrow Line Seyfert 1 galaxy (NLS1). The
detection of the L$\alpha$ line is possible here since the abundance
of iron is particularly strong.  The result is from a very long
(500~ks) XMM-Newton exposure which has also enabled the detection of
X-ray reverberation for the first time (Fig.~9, Fabian et al 2009).
This means that the reflection-dominated emission below 1~keV lags
behind the power-law which dominates the spectrum above 1~keV, owing
to the difference in light paths taken by the direct power-law and by
reflection (Fig.~2). The lag is about 30~s which corresponds to about
$2r_{\rm g}$ for a $2\times 10^6\Msun$ black hole, such as is
suspected in 1H\,0707-495.  The results imply that most of the primary
coronal X-ray source is very compact and centred close to the black
hole above the inner accretion disc.  We have now obtained similar lag
results from another NLS1, IRAS13224-3809 (also plotted in Fig.~9;
Fabian et al 2012).

\begin{figure}[h]
\begin{center}
\includegraphics[width=2.5in,angle=0]{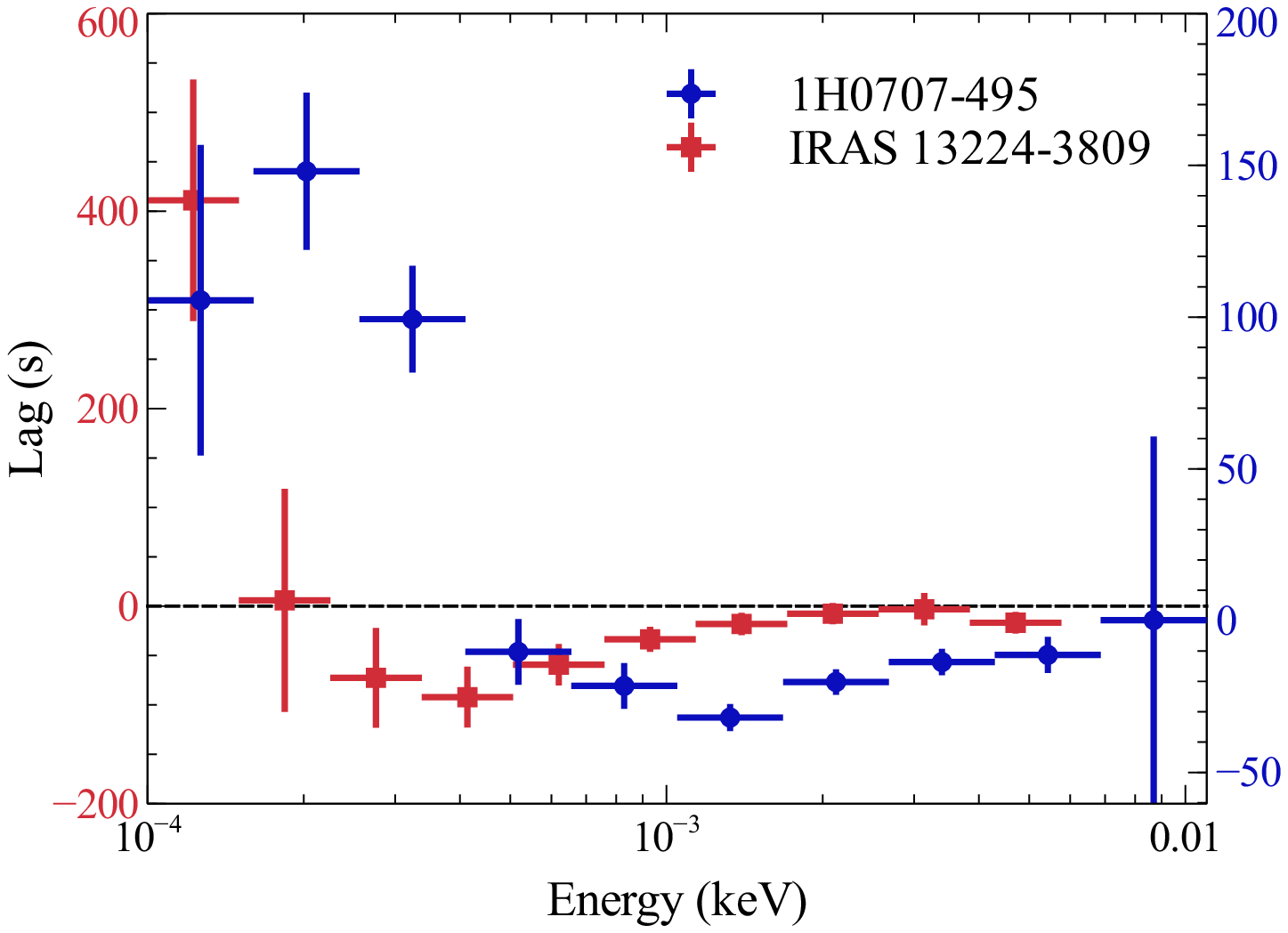}
\raisebox{4.5cm}{\includegraphics[width=1.9in,angle=-90]{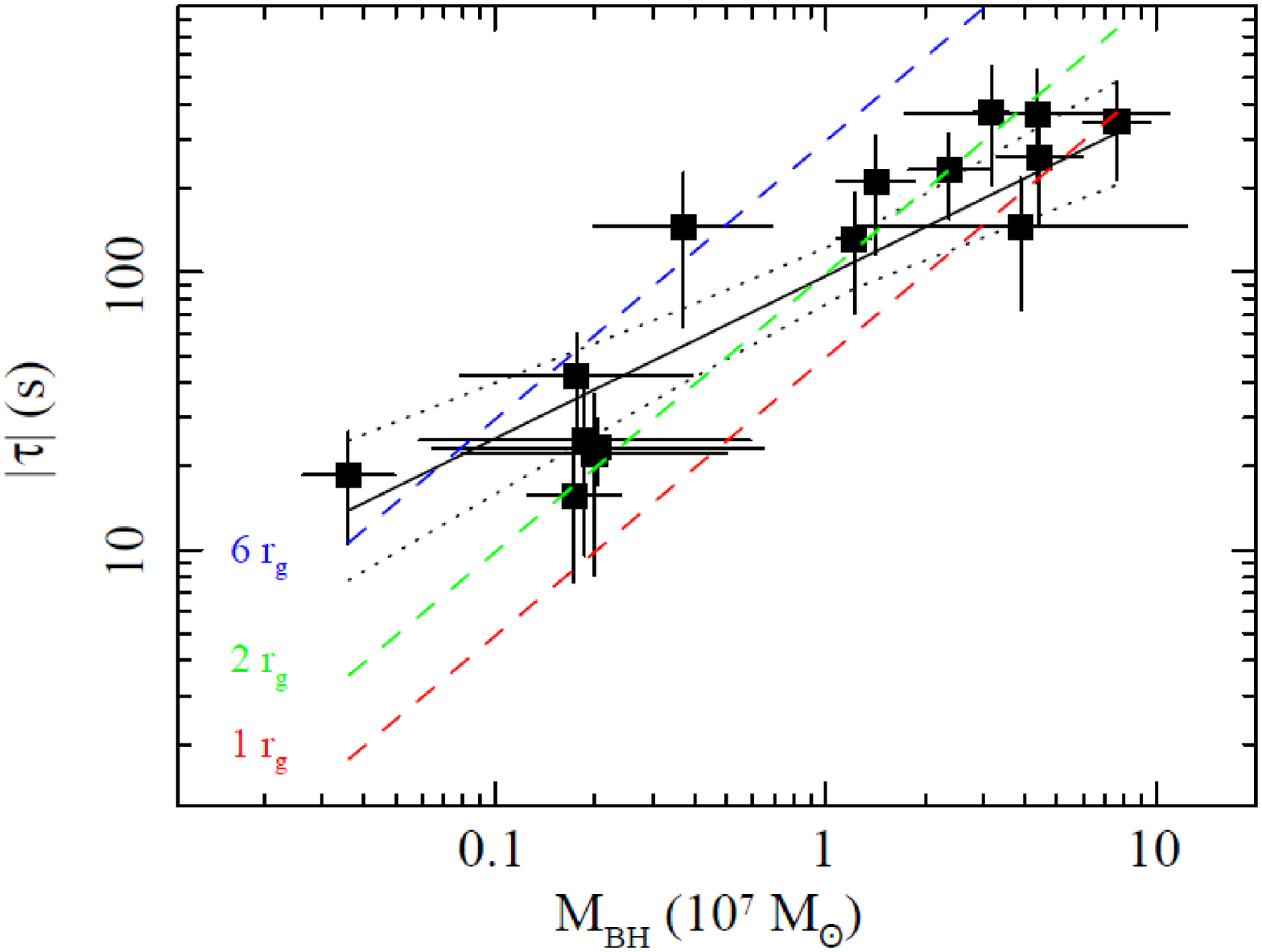}}
\end{center}
\caption{Top: Soft vs. Hard Lags in 1H\,0707-495 and IRAS13224-3809 as
  a function of frequency. At the highest frequencies the lag is
  negative meaning that changes in the reflection-dominated soft band
  lags behind those in the power-law-dominated hard band (Fabian et al
  2009; 2012).  Right: Timescale of soft lags (15 detected out of 32
  sources examined) plotted versus black hole mass (De Marco et al
  2012). The light crossing time of 1, 2 and $6r_{\rm g}$ are shown by
  dsahed lines.  }
\end{figure}

\begin{figure}[h]
\begin{center}
\includegraphics[width=2.7in,angle=0]{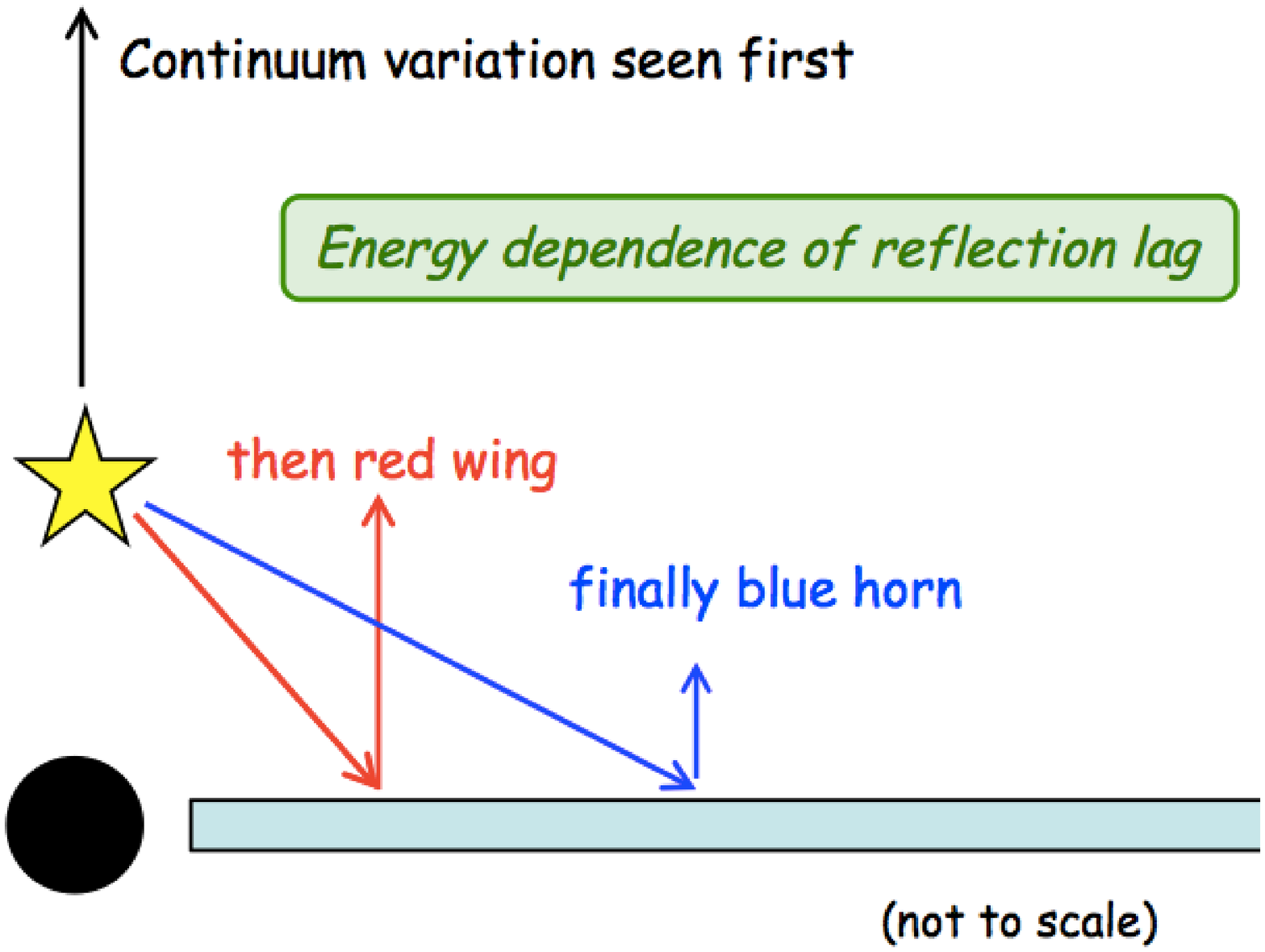}
\hspace{0.3cm}
\includegraphics[width=2.in,angle=0]{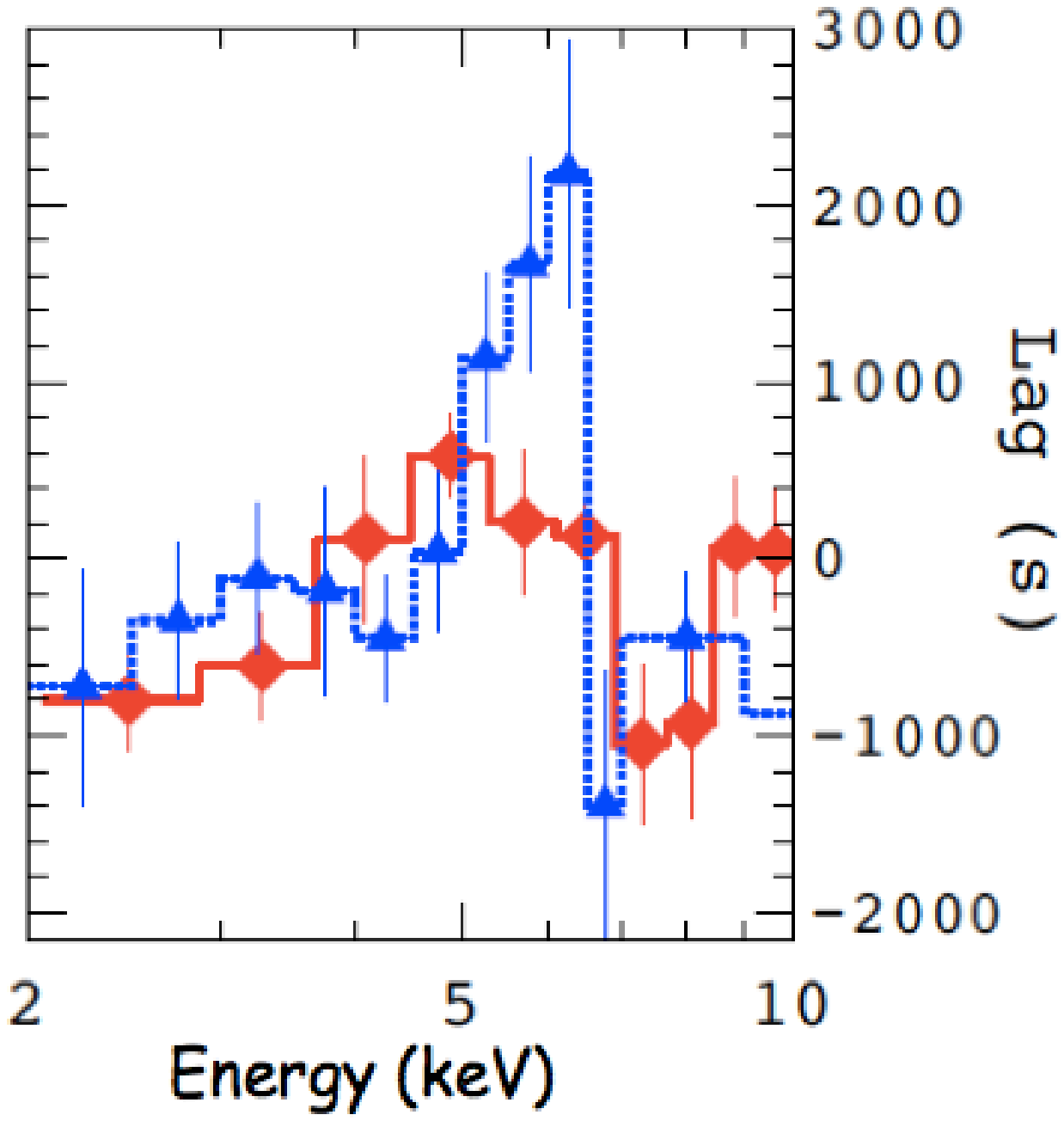}
\end{center}
\caption{Left: Schematic explanation of the energy dependence expected
  from reflection lags. A variation in the continuum is seen first
  then detected in the reflection spectrum from the innermost part of
  the disc so in the red wing of the iron line, and finally arrives
  from the outer part of the disc, seen in the blue horn.  Right: Lag
  energy spectra for NGC\,4151 (Zoghbi et al 2012). The shorter, high
  frequency lags (red) peak at 4--5~keV in the red wing of the lines,
  originating closest to the black hole, whereas the larger, lower
  frequency lags (blue) show a narrower spectral peak at 6--7~keV.  }
\end{figure}

A key test of the reverberation/reflection picture is the energy
dependence of the lags. Reflection occurs over a wide range of radii
in the disc. Variations in the primary continuum should be rapidly
followed by variations in the inner reflection, where the
gravitational redshift is strongest, so appearing in the red wing of
the broad line. The outer parts of the disc, giving the blue horn of
the line, respond slower and give larger lags (Fig.~10 left). This is
indeed what is observed in the first reverberation results from the
iron-K band obtained in the X-ray bright AGN, NGC\,4151 (Fig.~10
right, Zoghbi et al 2012). Iron K$\alpha$ lags are clearly seen with
the most rapid (higher temporal frequency) variations showing shorter
lags ($\sim 1000\s$) at more redshifted energies (4--5~keV) than the
slower (lower temporal frequency) variations showing larger lags
($\sim 2500\s$) close to the rest energy of 6--7~keV. Iron K$\alpha$
lags are now also seen in the in the deepest observations of 1H\,0707-495 and
IRAS13224-3809 (Kara et al 2012a,b).

A study of 32 AGN by Barbara De Marco et al (2012) lists 15
more AGN (9 above a significance of 3$\sigma$) showing rapid
reverberation in soft X-rays (Fig.~9). The lag timescales correlate
with mass and show that most of the reverberation originates from
within a few gravitational radii. This naturally leads to the
conclusion that soft X-ray emission from many AGN originates close to
the ISCO around moderately to highly spinning black holes. Finally,
the similarity between the geometry of AGN and BHB is emphasised by
the discovery of ms timescale lags in the BHB  by Uttley et al (2011). 

Note that there is a bias in any flux-limited sample of AGN towards
highly-spinning objects if the distribution of mass accretion rate at
large radii is the same for all spins (Brenneman et al 2011). The mass
to radiation conversion efficiency of an accretion disc increases by a
factor of 3 or more as the spin is increased.

A detailed discussion of the energy and frequency development of the
lags, including the Shapiro delay, is given by Wilkins \& Fabian (2012b).

\section{Summary}

We now have very good observational evidence of the strong gravity
regime around black holes in AGN and BHB. X-ray observations enable us
to probe General Relativistic effects such as large gravitational
redshifts, strong light bending and an ISCO at radii implying dragging
of inertial frames. The overall picture obtained in this way is
consistent with GR but does not yet test it. That may come about
through testing models in which GR is modified (e.g. Johanssen \&
Psaltis 2012) or, possibly, through a combination of more precise
light bending (i.e. space) and reverberation (time) measurements.

\section{Acknowledgements}
I am grateful to the Conference Organisers for the opportunity to talk
at this interesting meeting. Thanks to my many collaborators,
including Dan Wilkins, Erin Kara, Dom Walton, Abdu Zoghbi, Rubens
Reis, Phil Uttley, Ed Cackett, Jon Miller, Luigi Gallo, Giovanni
Miniutti, Chris Reynolds and Randy Ross, and to George Chartas for
Fig.~1 (right).

\end{document}